\documentclass[12pt,a4paper]{article}
\pdfoutput=1
\usepackage{amssymb,amsmath,amsfonts, mathtools, mathrsfs, graphicx, caption, subcaption, doi, cite, hyperref, url, dsfont}
\usepackage[utf8]{inputenc} 
\usepackage[dvipsnames]{xcolor}

\usepackage{verbatim}
\usepackage{amsfonts,euscript,amssymb,stmaryrd,braket}
\usepackage{tikz}
\usetikzlibrary{arrows,decorations.markings,patterns}
\usepackage{slashed}

\def\clock{{\count0=\time
		\divide\count0 60
		\ifnum\count0<10 0\fi\the\count0
		\multiply\count0 -60 \advance\count0 \time
		:\ifnum\count0<10 0\fi \the\count0
}}
\newcommand{\timestamp}{{\small\vbox{\hbox{\tt\jobname.tex}
			\hbox{\the\day/\the\month/\the\year, \clock}}}}

\makeatletter
\let\old@startsection=\@startsection
\let\oldl@section=\l@section
\renewcommand{\@startsection}[6]{\old@startsection{#1}{#2}{#3}{#4}{#5}{#6\mathversion{bold}}}
\renewcommand{\l@section}[2]{\oldl@section{\mathversion{bold}#1}{#2}}
\makeatother

\numberwithin{equation}{section}

\begin{document}
	\renewcommand{\thefootnote}{\arabic{footnote}}

	\overfullrule=0pt
	\parskip=2pt
	\parindent=12pt
	\headheight=0in \headsep=0in \topmargin=0in \oddsidemargin=0in

	\vspace{ -3cm} \thispagestyle{empty} \vspace{-1cm}
	\begin{flushright} 
		\footnotesize
		\textcolor{red}{\phantom{print-report}}
	\end{flushright}

\begin{center}
	\vspace{1.2cm}
	{\Large\bf \mathversion{bold}
	Entanglement from holography for five-dimensional charged Schr\"{o}dinger black hole}
	\\
	\vspace{.2cm}
	\noindent
	{\Large\bf \mathversion{bold}}

	\vspace{0.4cm} {
		Suman Das$^{\,a,}$\footnote[1]{suman.das@saha.ac.in} and Sabyasachi Maulik$^{\,b,}$\footnote[2]{mauliks@iitk.ac.in}},
	\vskip  0.4cm
	
	\small
	{\em
		$^{a}\,$Theory division, Saha Institute of Nuclear Physics,\\ A CI of Homi Bhabha National Institute, 1/AF, Bidhannagar, West Bengal 700064, India.
		\vskip 0.1cm
		$^{b}\,$Department of Physics, Indian Institute of Technology Kanpur,\\ Kalyanpur, Kanpur, Uttar Pradesh 208016, India.
	}
	\normalsize
	
\end{center}

\vspace{0.1cm}
\begin{abstract} 
We study entanglement entropy in a non-relativistic Schr\"{o}dinger field theory at finite temperature and electric charge using the principle of gauge/gravity duality. The spacetime geometry is obtained from a charged AdS black hole by a null Melvin twist. By using an appropriate modification of the holographic Ryu-Takayanagi formula, we calculate the entanglement entropy, mutual information, and entanglement wedge cross-section for the simplest strip subsystem. The entanglement measures show non-trivial dependence on the black hole parameters.
\end{abstract}
\section{Introduction} \label{intro}

Quantum entanglement happens to be a primary feature of quantum mechanics. A quantitative measure of quantum entanglement between the constituent parts of a bipartite quantum system is the entanglement entropy. The latter is a very useful quantity in many-body physics because it acts as an order parameter in quantum phase transitions \cite{Vidal:2002rm, Latorre:2003kg}, and also helps probe non-equilibrium processes such as a quantum quench \cite{Calabrese:2005in, Calabrese:2007mtj}. Hence, there is an active interest in studying entanglement entropy from a multitude of perspectives. Recent progress in experimental measurement of entanglement entropy \cite{Islam:2015mom} also adds a stimulus to this line of research. 

There are few theoretical tools available for an analytical calculation of entanglement entropy in a continuum quantum field theory. The most trustworthy methods work for low-dimensional theories $\left(D \leq 2 \right)$ with conformal symmetry \cite{Holzhey:1994we, Calabrese:2004eu, Calabrese:2009qy}, and they do uncover some important universal attributes. Nevertheless, it is useful to build a toolbox that aids computation for more general systems. For higher dimensional $\left(D \geq 3 \right)$ conformal field theories, the AdS/CFT correspondence \cite{Maldacena:1997re, Gubser:1998bc, Witten:1998qj}, or holography provides a means to achieve this goal. AdS/CFT correspondence relates the entanglement entropy between a sub-region $A$ and its complement in a strongly-coupled quantum field theory to the area of an extremal hypersurface $\Gamma$ in the dual spacetime, which is homologous to $A$ via the Ryu-Takayanagi (RT) proposal \cite{Ryu:2006bv, Ryu:2006ef, Hubeny:2007xt}
\begin{equation}\label{RT_formula}
	\mathcal{S}_{A} = \frac{\mathcal{A}_{\Gamma}}{4 G_N^{(D)}}, \quad \partial A = \partial \Gamma.
\end{equation}
Over the years, the holographic entanglement entropy (HEE) proposal has come to be regarded as one of the pioneering achievements of the AdS/CFT correspondence. 

Naturally, attempts have been made to generalize the AdS/CFT correspondence beyond its original incarnation in order to apply this powerful technique in physical systems more amenable to low-energy experiments. Of particular note are non-relativistic versions of gauge/gravity duality which attempt to describe strongly-coupled quantum field theories that govern physical systems relevant to condensed matter physics \cite{Son:2008ye, Balasubramanian:2008dm, Kachru:2008yh}. These systems could be ultracold atoms at unitarity which have the Schr\"{o}dinger symmetry \cite{Mehen:1999nd}, or a collection of strongly-coupled electrons at a critical point described by a Lifshitz field theory \cite{PhysRevLett.35.1678, PhysRevB.23.4615}. In this article we use a holographic description of the former with the motivation to extract some knowledge of quantum information in the context of non-relativistic field theories.

As the name suggests, a Schr\"{o}dinger field theory is a quantum field theory (QFT) which shares the symmetry group of the Schro\"{o}dinger equation in free space. They are non-relativistic analogues of conformal field theories (CFTs). The generators of the Schr\"{o}dinger algebra include the Hamiltonian $H$, linear momenta $P_i$, angular momenta $M_{ij}$, Galilean boost $K^i$, dilatation $D$, one special conformal transformation $C$, and the particle number (or mass) operator $N$ which commutes with all other generators. The non-relativistic dilatation and special conformal transformation operators act on the spacetime coordinates in the following way
\begin{align*}
    &D:\qquad t \rightarrow e^{2\lambda}t,\quad x^j \rightarrow e^{\lambda}x^j,\\
    &C:\qquad t \rightarrow \frac{t}{1+\lambda t},\quad x^j \rightarrow \frac{x^j}{1+\lambda t},
\end{align*}
the anisotropic scaling of time and space coordinates is a typical property of a non-relativistic field theory, and the relative scale dimension between time and space is called the `dynamical exponent' which for the present case is $z=2$ \footnote{it is possible to construct a theory with $z\neq 2$, but they would lack a special conformal transformation.}. The spacetime metric of a $\left(d+3\right)$ dimensional spacetime geometry which realizes the full Schr\"{o}dinger symmetry can be expressed as \cite{Son:2008ye, Balasubramanian:2008dm}
\begin{equation} \label{Sch_metric_gs}
    ds^2 = -\frac{dt^2}{r^4} + \frac{2\,dt\, d\xi}{r^2} + \frac{dr^2 + dx^{i}dx_{i}}{r^2}, \quad \left(i = 1, 2, \ldots, d\right).
\end{equation}
The usual AdS/CFT dictionary can be used on this spacetime background to calculate correlation functions of a $\left(d+1\right)$ dimensional non-relativistic conformal field theory (NRCFT) \cite{Balasubramanian:2008dm}. This lead to the hypothesis that the spacetime \eqref{Sch_metric_gs} is the gravity dual of the ground state of a Schr\"{o}dinger theory. It is worth mentioning that unlike the usual gauge/gravity duality where the bulk theory has one extra spatial dimension, here we encounter a codimension 2 correspondence. In addition to the radial direction $r$, the spacetime geometry has another noncompact dimensions $\xi$. On the boundary, translation along the $\xi$ coordinate is mapped to the mass or particle number operator $N = i\partial_{\xi}$. Care should be taken while doing calculations using Schro\"{o}dinger holography because particle number remains conserved in non-relativistic theories .

Finite temperature and finite density solutions with non-relativistic asymptotics were constructed in \cite{Herzog:2008wg, Maldacena:2008wh, Adams:2008wt, Imeroni:2009cs,Adams:2009dm}. In this article, we consider the five dimensional charged Schr\"{o}dinger black hole found in \cite{Adams:2009dm}. We want to use the Ryu-Takayanagi formula \eqref{RT_formula} adapted to this spacetime to calculate the holographic entanglement entropy, mutual information, and entanglement wedge cross-section for the simplest strip-shaped subsystem in the boundary NRCFT. The charged Schr\"{o}dinger black hole is obtained from a five dimensional Reissner-Nordstr\"{o}m AdS black hole by a null Melvin twist \cite{Gimon:2003xk, Alishahiha:2003ru}. It was shown in the original paper \cite{Adams:2009dm} that the non-relativistic background inherits some of the features of the parent AdS geometry, including the thermodynamic entropy. We want to see whether the entanglement structure is also inherited directly or whether it depends explicitly on the Melvinization. We discover that it is the latter. The three information theoretic quantitites studied in this paper are observed to depend on the parameters of the underlying theory in a manner quite different from one another. Entanglement in Schr\"{o}dinger field theory has been studied before, both using holographic \cite{Kim:2012nb, Roychowdhury:2016wca} and field-theory technique \cite{Mintchev:2022xqh, Mintchev:2022yuo}. 

The rest of the article is organized as follows: we start by introducing the charged Schr\"{o}dinger black hole geometry in section \ref{sec2}. In section \ref{sec3} we study the holographic entanglement entropy in this geometry for subsystems which take the shape of a long strip. Besides entanglement entropy, we also comment on the mutual information. In section \ref{sec4} we study the entanglement wedge cross-section for very thin strip. We end in section \ref{conclusions} with some remarks.

\section{Charged Schr\"{o}dinger black hole} \label{sec2}

It is possible to find a spacetime which realizes the Schr\"{o}dinger group as its asymptotic symmetry group by starting from an asymptotically AdS spacetime, and applying a series of symmetry transformations. This operation is known as the null Melvin twist \cite{Gimon:2003xk, Alishahiha:2003ru} and also as a TsT transformation. The procedure has been successfully applied to AdS black hole in type-IIB supergravity to obtain new black hole solutions with asymptotically Schr\"{o}dinger geometry \cite{Maldacena:2008wh, Herzog:2008wg, Adams:2008wt, Imeroni:2009cs, Adams:2009dm} both with and without charge and/or rotation. In this paper, we work with a five dimensional charged Schr\"{o}dinger black hole constructed in \cite{Adams:2009dm} by using the null Melvin twist on a Reissner-Nordstr\"{o}m AdS black hole. The result is a static spacetime dual to a non-relativistic quantum field theory at finite temperature and charge density.

The five dimensional charged Schr\"{o}dinger solution in Einstein frame is written as
\begin{align} \label{metric}
	\begin{split}
	ds^2 &= \frac{K(r)^{\frac{1}{3}}}{r^2}\left[-\frac{f(r)}{H(r)^2 K(r)}d\tau^2 - \frac{f(r)\beta^2}{r^2 H(r) K(r)}\left(d\tau+dy\right)^2 + \frac{H(r)}{K(r)}dy^2 \right.\\ & \hspace{7.5 cm} \left. + H(r)\left(\frac{dr^2}{f(r)}+dx_1^2+dx_2^2 \right) \right],
	\end{split}\\
    A_q &= \frac{q r^2}{H(r) r_H^2}\left(1 + q^2 r_H^2\right)^{\frac{3}{2}} d\tau,
\end{align}
where
\begin{equation}
	H(r) = 1+q^2r^2, \hspace{2.5 mm} f(r) = H(r)^3 - \frac{r^4}{r_H^4}\left(1 + q^2r_H^2\right)^3, \hspace{2.5 mm} K(r) = 1 + \frac{\beta^2r^2\left(1 + q^2r_H^2\right)^3}{H^2r_H^4}.
\end{equation}
Here $q$ and $r_H$ are respectively the charge of the black hole, and the radius of its (outer) horizon \footnote{Our notation differs slightly from the original reference \cite{Adams:2009dm}, where the horizon of the pre-Melvinization AdS black hole enters the spacetime metric as a parameter, while the true outer horizon is denoted by $r_{+}$.}. The constant $\beta$ is related to the null Melvin twist. The gauge field $A_q$ is directly inherited from the parent Reissner-Nordstr\"{o}m AdS black hole.\\

To write the spacetime geometry in a more familiar form, we transform to light-cone coordinates $\left(\tau, y\right) \rightarrow \left(t, \xi\right)$
\begin{equation} \label{lcone_coord}
	t = \beta\left(\tau + y\right), \quad \xi = \frac{1}{2 \beta}\left(y- \tau\right),
\end{equation}
the charged Schr\"{o}dinger black hole geometry \eqref{metric} in light cone coordinates \eqref{lcone_coord} is described by the metric
\begin{equation} \label{metric_ADM}
		ds^2 = \alpha^2(r) d\xi^2 + \sigma(r)\left(\beta(r)d\xi + dt\right)^2 + \frac{H(r) K(r)^{\frac{1}{3}}}{r^2}\left(\frac{dr^2}{f(r)} + dx_1^2 + dx_2^2\right),
\end{equation}
where
\begin{align}
	\alpha^2(r) &= -\frac{4 \beta ^2 f(r) \left(\beta ^2 \left(H(r)^3-f(r)\right)+r^2 H(r)^2\right)}{r^2 K(r)^{2/3} \left(r^2 H(r)^4-f(r) H(r) \left(4 \beta ^2 H(r)+r^2\right)\right)},\\
	\beta(r) &= \frac{2 \beta ^2 r^2 \left(f(r)+H(r)^3\right)}{r^2 H(r)^3-f(r) \left(4 \beta ^2 H(r)+r^2\right)},\\
	\sigma(r) &= \frac{r^2 H(r)^3-f(r) \left(4 \beta ^2 H(r)+r^2\right)}{4 \beta ^2 r^4 H(r)^2 K(r)^{2/3}}.
\end{align}

By setting $q=0$ and $r_H\to \infty$ we obtain the spacetime
\begin{equation}
	ds^2 = -\frac{dt^2}{r^4} + \frac{2\,dt\, d\xi}{r^2} + \frac{dr^2 + dx_1^2 + dx_2^2}{r^2},
\end{equation}
dual to the ground state of a zero-temperature three-dimensional quantum field theory with Schr\"{o}dinger symmetry \cite{Son:2008ye}.

Thermodynamics of the charged black hole in equation \eqref{metric} was extensively discussed in \cite{Adams:2009dm}. For instance, the thermodynamic entropy and temperature are
\begin{equation}
    S_{\mathrm{Th}} = \frac{H\left( r_H \right)^{\frac{3}{2}}}{4 G_N^{(5)} r_H^3}, \quad T_{\mathrm{Th}} = \frac{\sqrt{H\left(r_H\right)}}{2\pi\beta r_H} \lvert H\left(r_H\right) - 3\rvert,
\end{equation}
on the other hand, the parameter $\beta$ gets related to the chemical potential for $\xi$-translation
\begin{equation} \label{eqn:chempot}
    \mu_{\xi} = - \frac{1}{2\beta^2}.
\end{equation}
The thermodynamic entropy is manifestly independent of $\beta$, and hence does not change by the null Melvin twist. We will see shortly that this statement does not apply anymore for the entanglement entropy. The temperature of the black hole becomes zero when $H\left(r_H\right) = 3$; thus for $q r_H = \sqrt{2}$, the black hole becomes extremal.

\section{Entanglement entropy for a strip} \label{sec3}

We are interested in the entanglement between a sub-region $A$ and its complement on the conformal boundary $\left(r\to 0 \right)$ of the Schr\"{o}dinger spacetime \eqref{metric_ADM}. Since the spacetime is static, it is convenient to work on a constant time-slice of the geometry. We consider that the sub-region $A$ is of the shape of a strip defined by
\[A: \quad 0\leq \xi\leq L_{\xi}, \quad -\frac{\ell}{2}\leq x_1 \leq \frac{\ell}{2}, \quad 0\leq x_2\leq L \quad \text{such that } L\gg \ell. \]

As has already been stated in the introduction, Schr\"{o}dinger holography is a  co-dimension 2 correspondence. As such it is not evident if the holographic proposal for entanglement entropy can be applied in this context as it is. This puzzle was already addressed in \cite{Kim:2012nb}. The light-cone coordinate $\xi$ has a special role in this duality; translation along this direction is identified with the conserved mass or particle number operator $N = i\partial_{\xi}$ in the Schr\"{o}dinger algebra, which commutes with all other operators. The prescription in \cite{Kim:2012nb} recommends that one should let the $\xi$ coordinate span some finite range \[0\leq\xi\leq L_{\xi}\] in the gravitational point-of-view, and integrate over the entire $\xi$ coordinate while evaluating the minimal surface. From the ADM-like decomposition \eqref{metric_ADM} we see that the length along the $\xi$ direction is measured by $\alpha\left(r\right) d\xi$. We multiply this length element inside the integral in the area functional.

Due to the translation symmetry along $x_2$ direction, a bulk hypersurface $\Gamma$ homologous to the sub-region $A$ can be embedded by a single function $x_1=x(r)$, and its area is found by evaluating the integral
\begin{equation} \label{area_func}
	\mathcal{A}_{\Gamma} = 2 L L_{\xi} \int_{\epsilon}^{r_{\ast}} \alpha(r) \frac{H(r) K(r)^\frac{1}{3}}{r^2} \sqrt{\frac{1}{f(r)} + x'(r)^2}\,dr ,
\end{equation}
subject to the boundary conditions (i) $x'(r)\rvert_{r_*}\to \infty$, and (ii) $\lvert x\left(r=\epsilon\right)\rvert = \frac{\ell}{2}$. Here we use the standard notation $x'(r) = \frac{dx}{dr}$. We put a cutoff at $r=\epsilon$ near the boundary to regulate any ultraviolet divergence in the entanglement, and $r_*$ represents the \textit{turning point} of the hypersurface -- it is the highest point that the surface can reach inside the bulk spacetime.

In order to find the holographic entanglement entropy, we need to extremize the area functional in equation \eqref{area_func}. We first note that the integrand in \eqref{area_func} has no explicit dependence on $x(r)$, therefore, there exists a constant $C_1$ such that
\begin{align}
	& \frac{H(r) K(r)^{\frac{1}{3}} \alpha(r) x'(r)}{r^2 \sqrt{\frac{1}{f(r)} + x'(r)^2}} = C_1, \nonumber \\
	\text{or,}\quad & x'(r)^2 = \frac{\frac{C_1^2\,r^4}{f(r)}}{H(r)^2 K(r)^{\frac{2}{3}} \alpha(r)^2 - C_1^2\,r^4},
\end{align}
the boundary conditions help determine the constant $C_1$ in terms of the turning point $r_{\ast}$
\begin{equation}
	C_1^2 = \frac{H\left(r_*\right)^2 K\left(r_{\ast}\right)^{\frac{2}{3}} \alpha\left(r_{\ast}\right)^2}{r_{\ast}^4}.
\end{equation}
Therefore,
\begin{align}
	\frac{dx}{dr} = \frac{\frac{r^2}{r_{\ast}^2}}{\sqrt{ f(r) \left( \frac{H\left(r\right)^2 K\left(r\right)^{\frac{2}{3}} \alpha\left(r\right)^2}{H\left(r_*\right)^2 K\left(r_{\ast}\right)^{\frac{2}{3}} \alpha\left(r_{\ast}\right)^2} - \frac{r^4}{r_*^4} \right)} }, \label{turnpt_int}
	%
\end{align}
which upon substitution in \eqref{area_func} yields
\begin{equation} \label{area_func_2}
    \mathcal{A}_{\Gamma} = 2 L L_{\xi} \int_{\epsilon}^{r_*} \frac{dr}{r^2} \sqrt{\frac{\frac{ \left( H\left(r\right)^2 K\left(r\right)^{\frac{2}{3}} \alpha\left(r\right)^2 \right)^2}{H\left(r_*\right)^2 K\left(r_{\ast}\right)^{\frac{2}{3}} \alpha\left(r_{\ast}\right)^2}}{f(r) \left( \frac{H\left(r\right)^2 K\left(r\right)^{\frac{2}{3}} \alpha\left(r\right)^2}{H\left(r_*\right)^2 K\left(r_{\ast}\right)^{\frac{2}{3}} \alpha\left(r_{\ast}\right)^2} - \frac{r^4}{r_*^4} \right)}}. 
\end{equation}

The evaluation of holographic entanglement entropy thus comes down to solving the two integrals in equations \eqref{turnpt_int} and \eqref{area_func_2}. We should solve equation \eqref{turnpt_int} to obtain a relationship between the turning point $\left(r_*\right)$ and the width $\left(\ell\right)$ of the strip. Then after carrying out the integration in equation \eqref{area_func_2}, we obtain the area of the extremal Ryu-Takayanagi hypersurface as a function of the strip-width. In practice, however, it is difficult to obtain an exact analytic expression for the area since the integrals are extremely involved. In the following, we obtain the holographic entanglement entropy in two ways. First by numerical methods, and second by adapting a perturbative approach for very small strip-width.

\subsection{Numerical results for holographic entanglement entropy}

We adopt the following strategy to numerically obtain the holographic entanglement entropy $\left(S_A\right)$ as a function of strip-width $\left(\ell\right)$
\begin{enumerate}
    \item We create an array of possible values of the turning-point $\left( r_{\ast} \right)$.
    \item By numerically integrating equation \eqref{turnpt_int} for each $r_{\ast}$ value in the above array, we obtain an array of sub-region size $\left(\ell\right)$.
    \item We also numerically integrate equation \eqref{area_func_2} for each entry in the array of $r_{\ast}$ values with appropriate regularization to obtain a list of minimum area.
    \item Finally we plot the lists in (2) and (3) against each other to display the change in HEE with changing strip-width.
\end{enumerate}
\begin{figure}[t]
    \begin{subfigure}{0.48\textwidth}
        \centering
        \includegraphics[width=\textwidth]{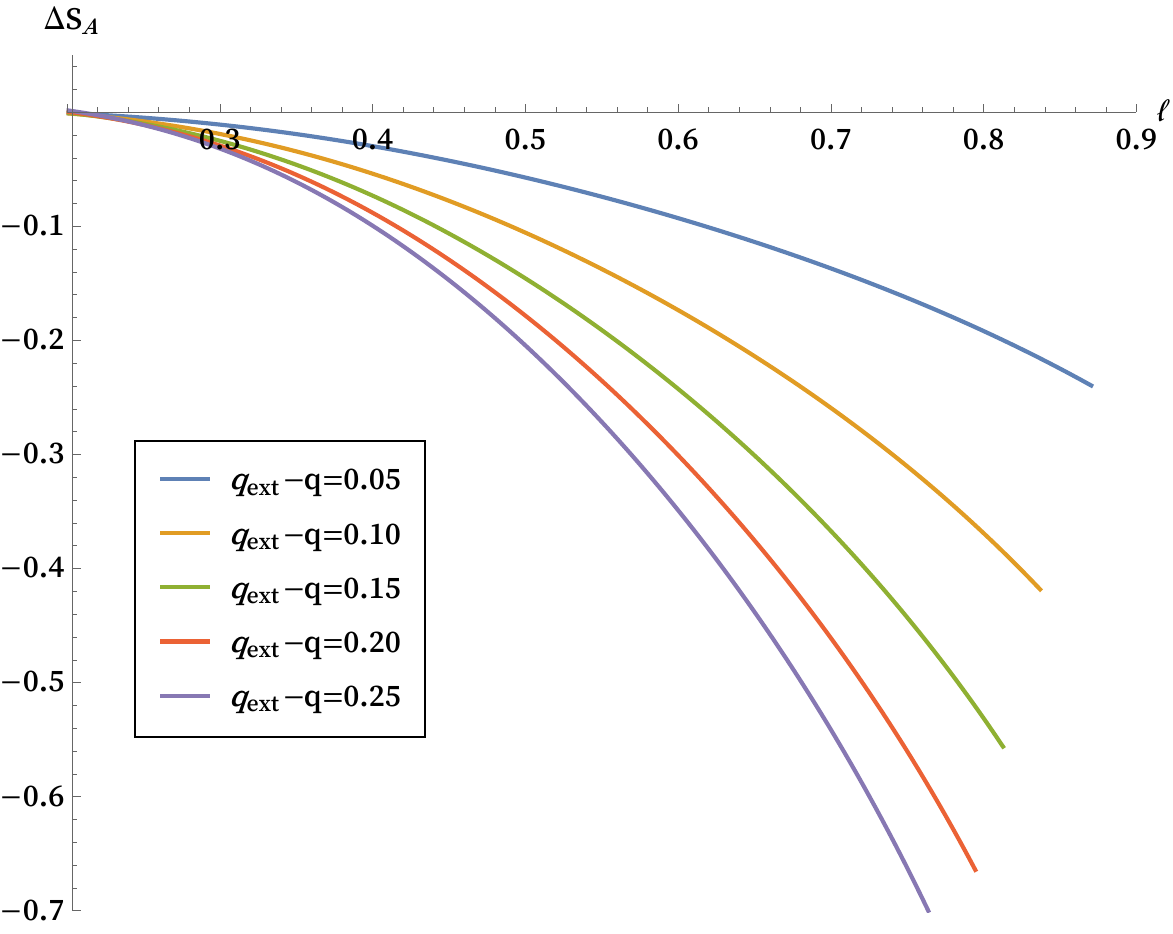}
        \caption{$\beta=0.75$}
    \end{subfigure}
    \hfill
    \begin{subfigure}{0.48\textwidth}
        \centering
        \includegraphics[width=\textwidth]{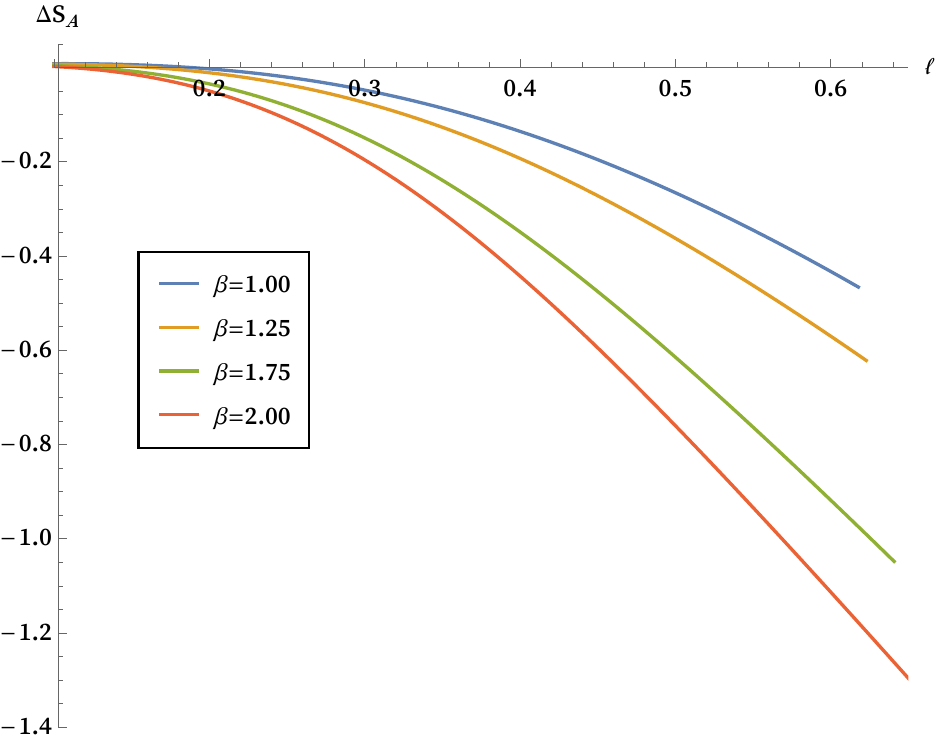}
        \caption{$q=0.75$}
        \label{fig_HEE_b}
    \end{subfigure}
    \caption{Change in HEE over the extremal black hole: (a) for different charge $q$, and (b) for different $\beta$ values. Plots drawn by choosing $2 L L_{\xi} = 1$, $r_H = \sqrt{2}$, and in units where $4 G_N^{(5)} = 1$.}
    \label{fig_HEE}
\end{figure}
The results of this numerical analysis are summarised in the two plots in figure \ref{fig_HEE}. The holographic entanglement entropy suffers from a characteristic small-distance divergence. To get rid of the divergent part, we subtract from our result the HEE for the extremal black hole $\left( T_{\mathrm{ext}} = 0, q_{\mathrm{ext}} = \frac{\sqrt{2}}{r_H} \right)$. The holographic dual state to the extremal black hole is the ground state of the boundary quantum field theory. Hence, essentially we are studying the change in HEE over the ground state of the non-relativistic system, this is denoted by $\Delta S_A = S_A - S_A^{(\mathrm{ext})}$. If the electric charge $q < q_{\mathrm{ext}}$, then $\Delta S_A < 0$, i.e. the holographic entanglement entropy is smaller than that of the extremal black hole. We note that the entanglement entropy for a fixed strip-width usually increases as we increase the value of the charge $q$ from below towards the extremal value $q=q_{\mathrm{ext}}$. Thus increased electric charge results in more amount of entanglement between the degrees of freedom in $A$ and its complement. We also observe from figure \ref{fig_HEE_b} that the parameter $\beta$ does influence the entanglement entropy, which in fact decreases with increasing value of $\beta$. This is unlike the thermodynamic entropy of charged Schro\"{o}dinger black hole, which is manifestly independent of $\beta$ \cite{Adams:2009dm}. Since the chemical potential $\mu_{\xi}$ and $\beta$ are related to each other by \eqref{eqn:chempot}, we conclude that for a fixed charge, the HEE increases if $\mu_{\xi}$ decreases. If we regard the chemical potential $\mu_{\xi}$ as the energy cost of adding more particles in the theory, we may interpret this result by saying entanglement increases with decreasing chemical potential because it becomes easier to add more degrees of freedom in the system.

\subsection{Perturbative entanglement entropy for small strip-width}

In this section we derive an analytic expression for holographic entanglement entropy in the present theory in a restricted regime of the parameter space. The analytic result will be useful later when we calculate the mutual information and entanglement wedge cross-section -- for which our numerical methods no longer work.

We assume that the strip-width $\left(\ell\right)$ is considerably smaller than the horizon parameter $\left(r_H\right)$. Since $r_{\ast}\sim\ell$, this assumption is equivalent to saying that the R-T hypersurface doesn't penetrate very deep into the bulk. For the convenience of calculation, let us change to the variables\[y = \frac{r}{r_{\ast}}, \quad q = q\,r_H, \quad \beta = \frac{\beta}{r_H}.\] We do a perturbation series expansion of equation \eqref{turnpt_int} in powers of the dimensionless parameter $\frac{r_{\ast}}{r_H}$ to obtain
\begin{equation}
	\begin{split}
	\int_{0}^{\frac{\ell}{2}} dx = r_{\ast} \int_{0}^{1} dy \frac{y^2}{\sqrt{1-y^4}} \left( 1 + \left(\frac{r_*}{r_H}\right)^2 \frac{q^2 \left( 2 - 3 y^2 - 3 y^4 \right) + \beta^2 \left( 1 + q^2 \right)^3}{2 \left(1+y^2\right)} \right).
	\end{split}
\end{equation}
Note that in order obtain a consistent series expansion we also have to assume $ q, \beta \ll 1$. The integrations can now be performed very easily
\begin{equation}
	\frac{\ell}{2} = r_{\ast} \frac{\Gamma\left(\frac{3}{4}\right)^2}{\sqrt{2 \pi}} \left( 1 - \frac{1}{2} \left( \frac{r_*}{r_H} \right)^2 q^2 - \left( \frac{r_*}{r_H} \right)^2 \beta^2 \left(1 + q^2\right)^3 \frac{1 - \frac{\Gamma\left(\frac{1}{4}\right)^2}{4 \Gamma\left(\frac{3}{4}\right)^2}}{4} \right).
\end{equation}
Let us define 
\begin{align} \label{constant_defn}
    b_0 = \frac{\Gamma\left(\frac{3}{4}\right)^2}{\sqrt{2 \pi}}, \quad b_1 = \frac{\Gamma\left(\frac{1}{4}\right)^2}{4 \sqrt{2\pi}},
\end{align}
such that
\begin{equation}
    \frac{\ell}{2} = r_{\ast} b_0 \left( 1 - \frac{1}{2} \left( \frac{r_*}{r_H} \right)^2 q^2 - \frac{1}{4} \left( \frac{r_*}{r_H} \right)^2 \beta^2 \left(1 + q^2\right)^3 \left(1 - \frac{b_1}{b_0} \right) \right).
\end{equation}
The last equation maybe inverted to express $r_*$ in terms of $\ell$ $\left(\text{up to leading order in }\frac{\ell}{r_H}\right)$
\begin{equation} \label{turnpt_exp}
	r_* = \frac{\frac{\ell}{2 b_0}}{1 - \frac{1}{2}\left(\frac{\ell}{r_H}\right)^2 \frac{q^2}{4 b_0^2} - \frac{1}{4} \left(\frac{\ell}{r_H}\right)^2 \frac{\beta^2 \left(1 + q^2 \right)^3 \left( 1 - \frac{b_1}{b_0} \right)}{4 b_0^2} }.
\end{equation}
To calculate the area of the extremal hypersurface, we first substitute equation \eqref{turnpt_int} in \eqref{area_func}, and do a similar perturbative expansion
\begin{align}
	\mathcal{A}_{\Gamma} = & 2 L L_{\xi} \int_{\epsilon}^{r_*} \frac{dr}{r^2} \sqrt{\frac{\frac{ \left( H\left(r\right)^2 K\left(r\right)^{\frac{2}{3}} \alpha\left(r\right)^2 \right)^2}{H\left(r_*\right)^2 K\left(r_{\ast}\right)^{\frac{2}{3}} \alpha\left(r_{\ast}\right)^2}}{f(r) \left( \frac{H\left(r\right)^2 K\left(r\right)^{\frac{2}{3}} \alpha\left(r\right)^2}{H\left(r_*\right)^2 K\left(r_{\ast}\right)^{\frac{2}{3}} \alpha\left(r_{\ast}\right)^2} - \frac{r^4}{r_*^4} \right)}}, \nonumber \\
    \begin{split}
    = & \frac{2 L L_{\xi}}{r_*} \int_{\frac{\epsilon}{r_{\ast}}}^{1} \frac{dy}{y^2 \sqrt{1 - y^4}} \left(1 - \left(\frac{r_*}{r_H}\right)^2 \frac{q^2 y^2 \left(1-y^2 \right)}{2 \left(1+y^2 \right)} \right. \\ &\hspace{4.4cm}\left. + \left(\frac{r_*}{r_H}\right)^2 \frac{\beta^2 \left(1+q^2\right)^3 y^2\left(1+2y^2\right)}{2\left(1+y^2\right)} \right), 
    \end{split}
	\nonumber \\
    \begin{split}
	= & \frac{2 L L_{\xi}}{\epsilon} + \frac{2 L L_{\xi}}{r_*} \left(- \frac{\pi}{4 b_1} - \left(\frac{r_*}{r_H}\right)^2 \frac{q^2}{4} \frac{\sqrt{\pi b_0}}{\sqrt{b_1}} \right. \\ &\hspace{3.9cm}\left. + \left(\frac{r_*}{r_H}\right)^2 \frac{\beta^2 \left(1+q^2\right)^3}{8} \left( \frac{3\sqrt{\pi b_1}}{\sqrt{b_0}} - \frac{\sqrt{\pi b_0}}{ \sqrt{b_1}} \right)  \right).
    \end{split}
\end{align}
Finally, by using equation \eqref{turnpt_exp} we arrive at
\begin{equation} \label{eqn:HEE_pert_answer}
    \begin{split}
        \mathcal{A}_{\Gamma} = \frac{2 L L_{\xi}}{\epsilon} &- \frac{L L_{\xi}}{\ell} \frac{\pi b_0}{b_1} \left(1 + \frac{\ell^2}{r_H^2} \frac{q^2}{8} \left( \frac{2\sqrt{b_1}}{\pi b_0^3} - \frac{1}{b_0^2} \right) \right. \\ &\hspace{2.55cm} \left. + \frac{\ell^2}{r_H^2} \frac{\beta^2 \left( 1 + q^2 \right)^3}{16} \left(-\frac{1}{b_0^2} + \frac{b_1}{b_0^3} + \frac{2\sqrt{b_1}}{\sqrt{\pi b_0^3}} - \frac{6\sqrt{b_1^3}}{\sqrt{\pi b_0^5}} \right) \right),
    \end{split}
\end{equation}
at leading order in our perturbative approach. The holographic entanglement entropy maybe found using the R-T proposal \eqref{RT_formula}. The HEE for the ground state of the zero-temperature, zero-charge theory can also be read off from equation \eqref{eqn:HEE_pert_answer}
\begin{equation}
   4 G_N^{(5)} \times S_A \rvert_{T_{\text{Th}}=0=q} = \frac{2 L L_{\xi}}{\epsilon} - \frac{L L_{\xi}}{\ell} \frac{\pi b_0}{b_1},
\end{equation}
where we observe the UV divergence characteristic of entanglement entropy. The HEE also shows correct scaling with the sub-region size $\left(\ell\right)$ for non-relativistic theories.\\

\begin{figure}[t]
\centering
\includegraphics[scale=0.46]{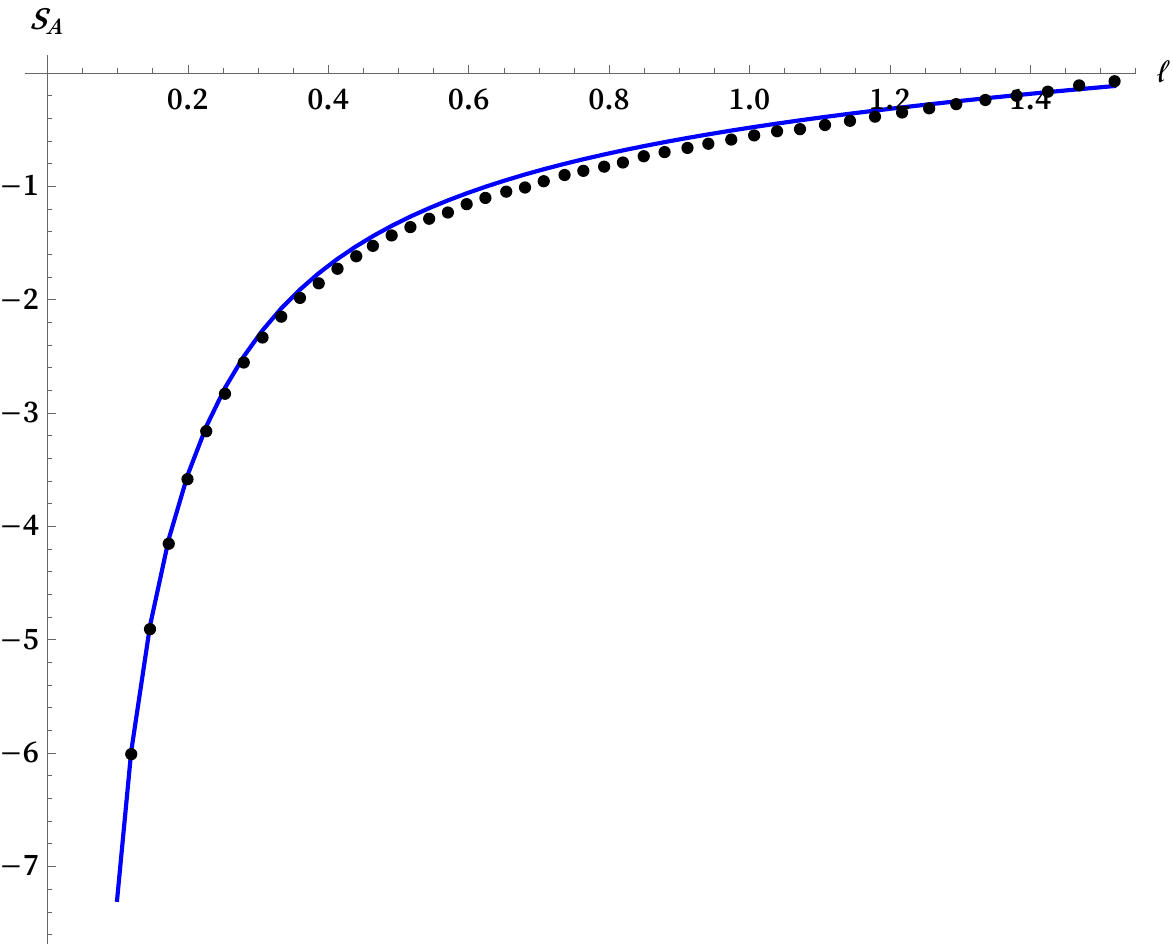}
\caption{Comparison of numerical and perturbative results for holographic entanglement entropy; plot is drawn in units where $4 G_N^{(5)} = 1$ by choosing $2 L L_{\xi} = 1$, and for the parameter values $r_H = 2.0$, $q = 0.25$, $\beta = 1.0$. The dotted curve denotes points obtained by numerics. The plot displays only the finite part of entanglement entropy.}
\label{fig:HEE_comparison}
\end{figure}

As a cross-check of the reliability of our perturbative computation, a comparison between the perturbative result \eqref{eqn:HEE_pert_answer} and the numerical result obtained in the previous section is portrayed in figure \ref{fig:HEE_comparison}. We observe that the perturbation series analysis is a very good approximation for small strip-width, and in the range of allowed parameter values.

\subsection{Mutual information}

\begin{figure}[h]
    \centering
    \includegraphics[scale=0.2]{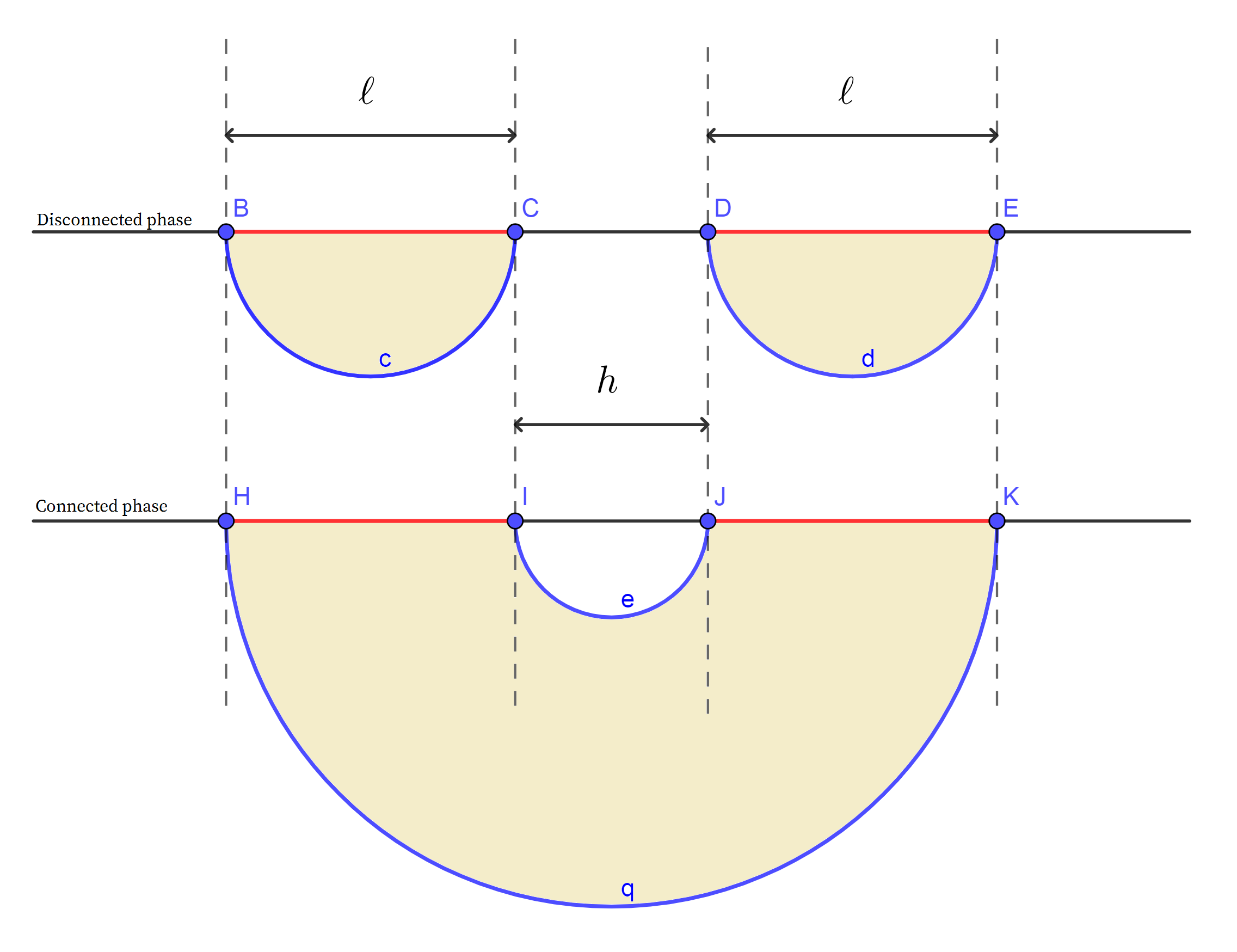}
    \caption{A symmetric configuration of two-strips of width $\ell$, and separation $h$. There are two competing R-T hypersurfaces for this configuration, and a geometric phase transition is observed from the `connected' to the `disconnected' phase when the separation between $A$ and $B$ is sufficiently large. The shaded region denotes the entanglement wedge in each phase.}
    \label{fig_HMI_setup}
\end{figure}

Entanglement entropy suffers from a small-distance divergence due to high amount of entanglement among the degrees of freedom lying near the boundary of the entangling surface. Hence, any calculation of entanglement entropy requires a regularization. One way to obtain a divergence-free quantity which is regularization independent is to construct appropriate linear combination of entanglement entropies. For a bipartite system, one popular such quantity is the mutual information
\begin{equation}
    I \left(A, B\right) = S_A + S_{B} - S_{A \cup B},
\end{equation}
where $A$ and $B$ are two disjoint portions of the sub-region situated on the spacetime background where the quantum field theory is defined (see figure \ref{fig_HMI_setup}). The mutual information $I\left(A, B\right)$ measures the total correlation between the regions $A$ and $B$.

To study mutual information in the present theory, we consider a subsystem $A\cup B$, where both $A$ and $B$ have widths $\ell$, and they are separated by a distance $h$. Due to the somewhat complicated metric function, our numerical methods are unable to yield appropriate answer. Hence, we use the approximate analytical result \eqref{eqn:HEE_pert_answer} to calculate $I\left(A, B\right)$ for different $q$ and $\beta$ values.

\begin{figure}[t]
    \begin{subfigure}{0.48\textwidth}
        \centering
        \includegraphics[width=\textwidth]{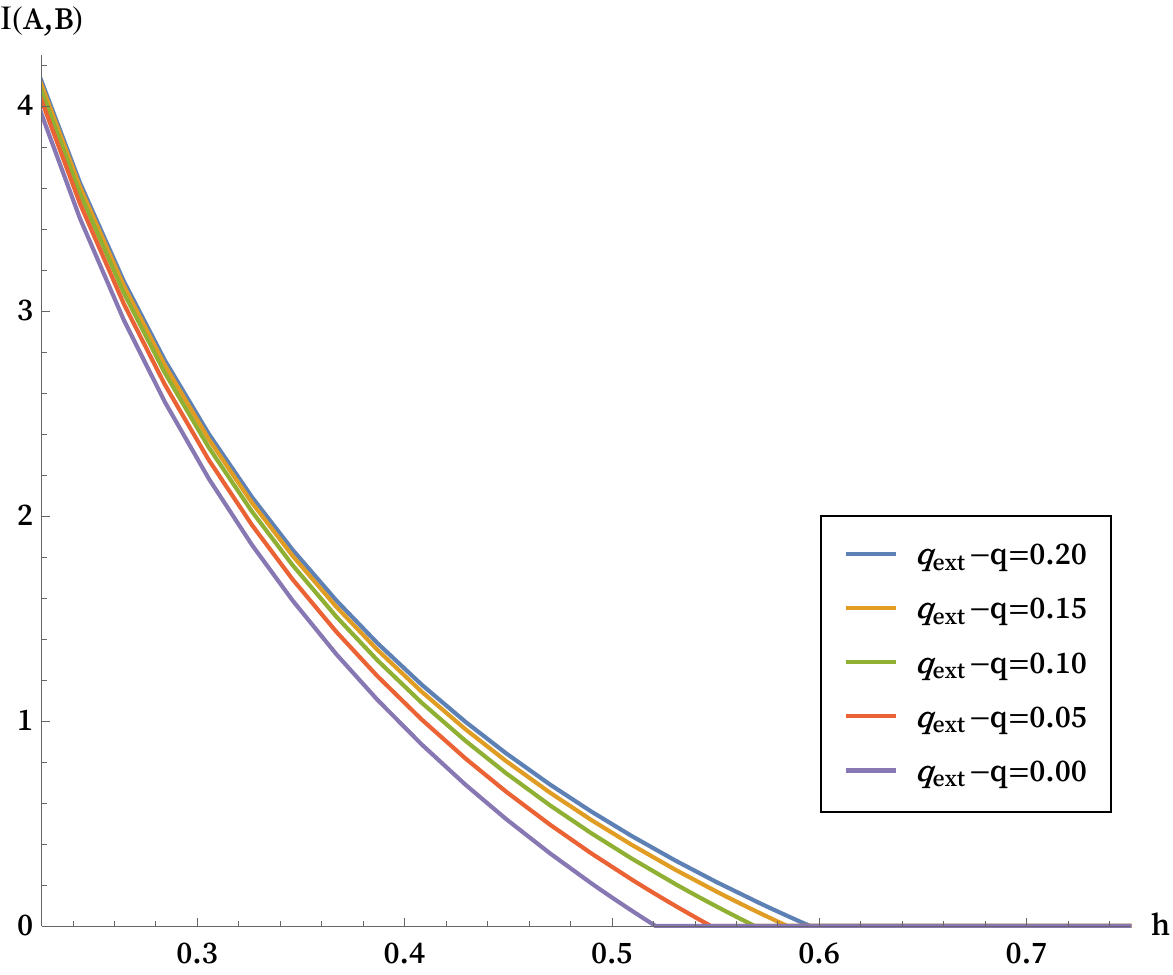}
        \caption{$\beta=1.0$}
    \end{subfigure}
    \hfill
    \begin{subfigure}{0.48\textwidth}
        \centering
        \includegraphics[width=\textwidth]{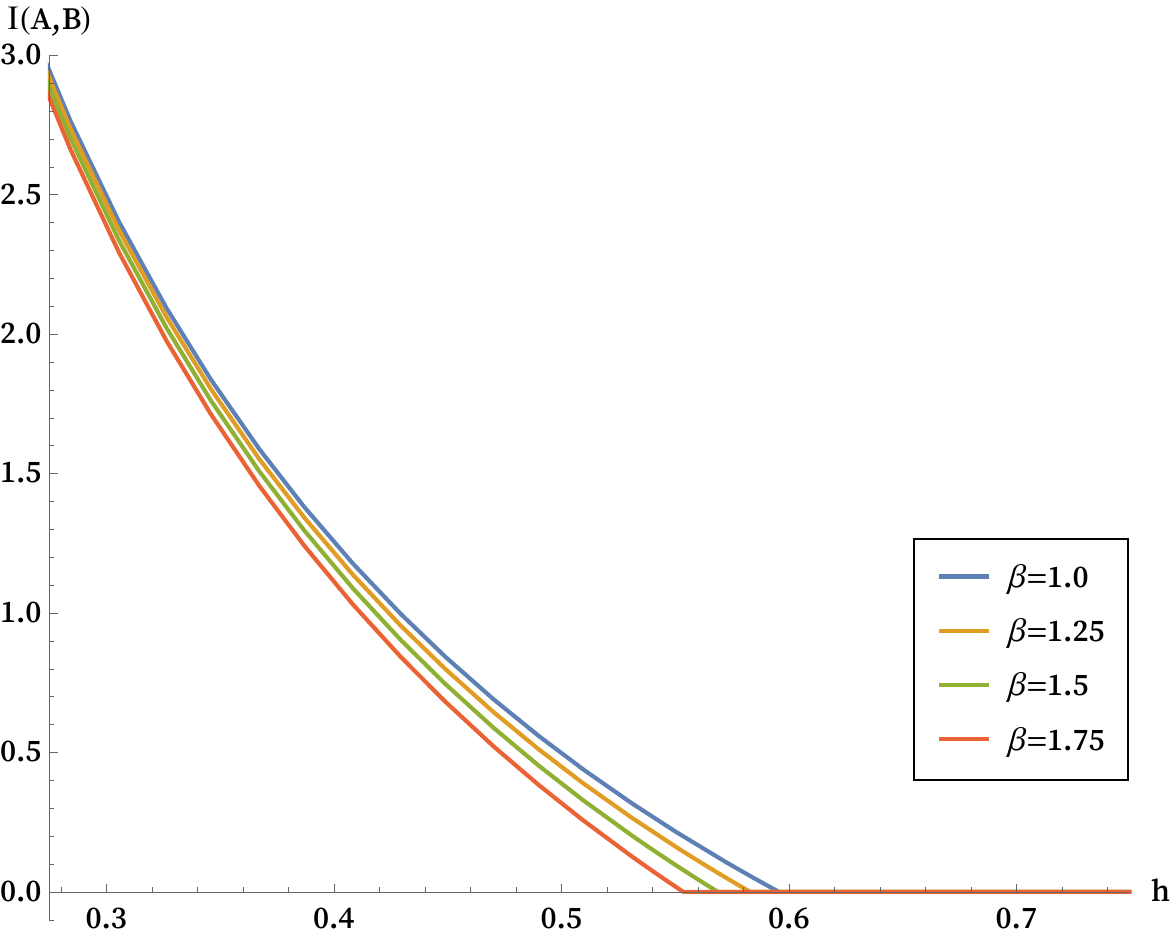}
        \caption{$q=0.30$}
    \end{subfigure}
    \caption{Dependence of mutual information $I\left(A, B\right)$ on the separation $h$ between $A$ and $B$: (a) for different charges $q$, and (b) for different $\beta$ values. In both cases $r_H = 2\sqrt{2}$, and $L=L_{\xi}=\ell=1$. Plots are drawn in $4G_N^{(5)}=1$ unit.}
    \label{fig_HMI}
\end{figure}

There are two possible extremal R-T hypersurfaces for a configuration of two disjoint regions. Depending on the separation between the two regions and their lengths; the mutual information undergoes a `first-order phase transition' from positive values to zero, see figure \ref{fig_HMI_setup}. This is because as the separation between $A$ and $B$ is increased, the hypersurface depicted in the upper part of figure \ref{fig_HMI_setup} becomes more dominant over the other. We can write
\begin{align} \label{MI_expression1}
	I\left(A:B\right) = \begin{dcases*}
		\mathcal{I}_{AB}, & if $ x < x_c $, \\
		0, & if $ x > x_c $,
	\end{dcases*}
\end{align}
where $x = \frac{h}{\ell}$, $x_c$ denotes the critical separation to strip-width ratio for which the phase transition takes place, and
\begin{equation}
\begin{split}
    4G_N^{(5)}\times\mathcal{I}_{AB} = -\frac{LL_{\xi}}{\ell}\frac{\pi b_0}{b_1} &\left(\left(2 - \frac{1}{x} - \frac{1}{x+2} \right) - \frac{\ell^2}{r_H^2} \left(x^2 + 2x + 1 \right) \left( \frac{q^2}{4} \left( \frac{2\sqrt{b_1}}{\pi b_0^3} - \frac{1}{b_0^2} \right) \right. \right. \\ &\left. \left. + \frac{\beta^2\left(1+q^2\right)^3}{8} \left(-\frac{1}{b_0^2} + \frac{b_1}{b_0^3} + \frac{2\sqrt{b_1}}{\sqrt{\pi b_0^3}} - \frac{6\sqrt{b_1^3}}{\sqrt{\pi b_0^5}} \right) \right) \right),
\end{split}
\end{equation}
We observe that for the Schr\"{o}dinger theory, the critical point usually decreases towards lower values of the separation length $h$ with increasing values of the charge $q$, and the non-zero value of the mutual information itself decreases with increasing $q$. This suggests that when the charge is increased, the total correlation between the degrees of freedom in region $A$ and $B$ decreases, and approaches zero faster. Similar features are observed when the charge is held fixed, but the other parameter $\beta$ is varied. It is seen that $I\left(A, B\right)$ decreases with increasing $\beta$ in the non-zero phase, and the `disentangling' phase transition occurs at a smaller separation length. See figure \ref{fig_HMI} for an illustration of these observations.

\section{Entanglement wedge cross-section} \label{sec4}

\begin{figure}[h]
    \centering
    \includegraphics[scale=0.2]{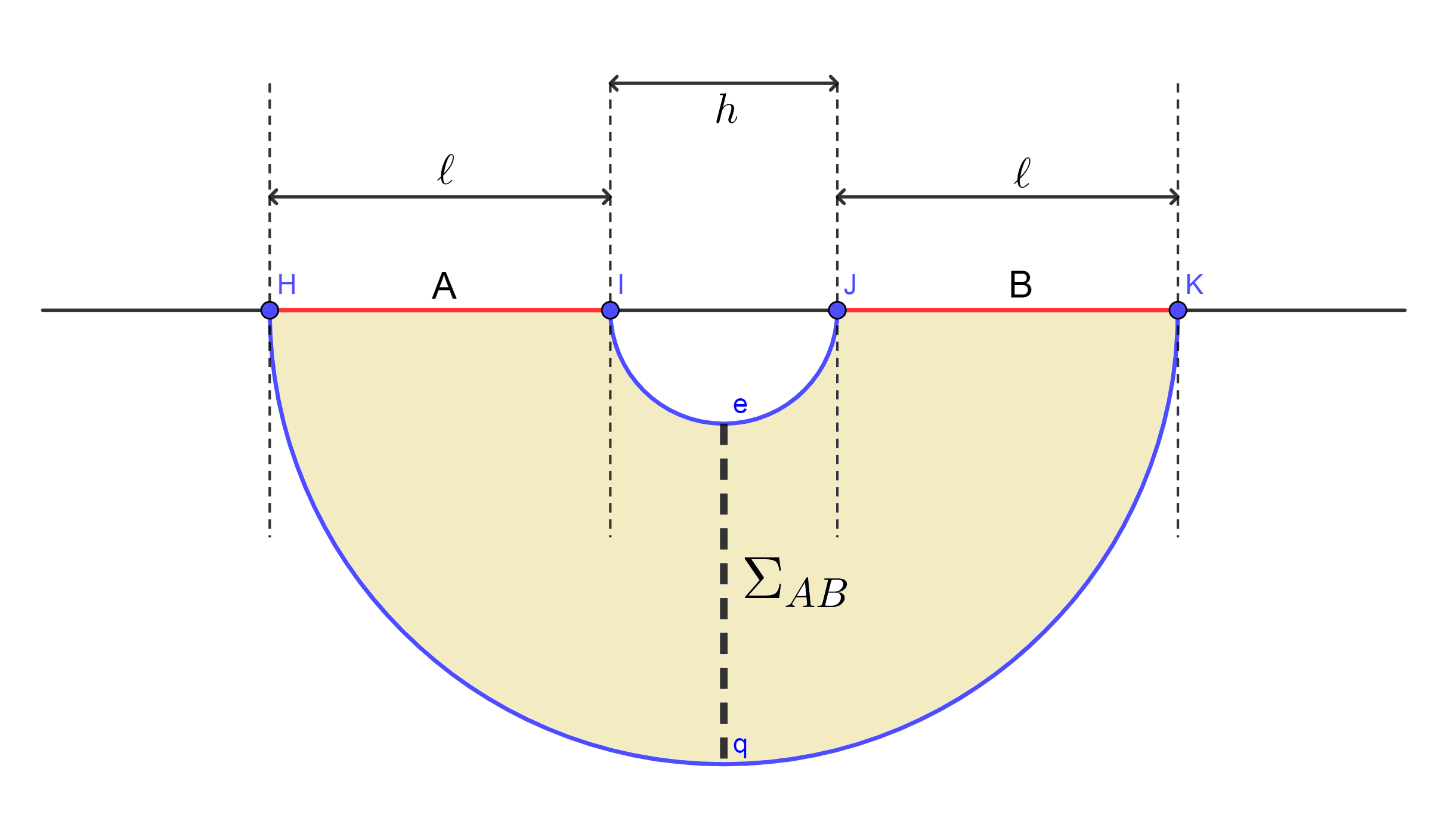}
    \caption{A symmetric configuration of two-strips of width $\ell$, and separation $h$. The entanglement wedge in the connected phase (when $I\left(A, B\right) \neq 0$) is the shaded region. The entanglement wedge cross-section is marked by the dashed line $\Sigma_{AB}$.}
    \label{fig_EWCS_setup}
\end{figure}

If the physical system under consideration is prepared in a mixed state, then entanglement entropy is no more the best measure of entanglement between the constituent parts. Therefore, new quantities are introduced which can measure the classical and quantum correlation (including entanglement entropy) between two subsystems. It is interesting to explore if they admit any dual interpretation in the context of holography. One such quantity is the mutual information $I\left(A, B\right)$ discussed in the previous section. However, being a linear combination of entanglement entropy, the mutual information is not a genuinely new quantity. In \cite{Takayanagi:2017knl}, it was reasoned that at least from the bulk geometric perspective there exists a natural candidate measure of the correlation between two disjoint subsystems. The part of the bulk spacetime which is considered dual to the reduced density matrix $\rho_{AB}$ of the boundary subsystem $A\cup B$ is called the entanglement wedge \cite{Wall:2012uf, Czech:2012bh, Headrick:2014cta}. Then the quantity
\begin{equation} \label{EWCS_defn}
	E_{W} \left(A, B\right) = \frac{\text{Area}\left(\Sigma_{AB}\right)}{4 G_N^{(5)}},
\end{equation}
where $\Sigma_{AB}$ refers to the minimum cross-section of the entanglement wedge (see figure \ref{fig_EWCS_setup}, can measure some correlation between $A$ and $B$. This quantity is called the entanglement wedge cross-section (EWCS). There is a lack of agreement on the field theoretic quantity which is the correct holographic dual to the EWCS. While the original papers proposed that the field-theoretic interpretation should be the entanglement of purification \cite{Takayanagi:2017knl, Nguyen:2017yqw}, later works showed that EWCS also shared resemblance with entanglement negativity \cite{Kudler-Flam:2018qjo, Kusuki:2019zsp}, odd entropy \cite{Tamaoka:2018ned}, and reflected entropy \cite{Dutta:2019gen}. The role of EWCS in holography for non-relativistic field theories has been studied from multiple perspectives in the fairly recent works \cite{BabaeiVelni:2019pkw, Gong:2020pse, Khoeini-Moghaddam:2020ymm, Vasli:2022kfu, BabaeiVelni:2023cge}.

The entanglement wedge cross-section also exhibits a geometric phase transition similar to mutual information. It is positive when the sub-regions $A$ and $B$ are close to each other. For sufficiently large separation between $A$ and $B$, the entanglement wedge is simply a union of two disconnected bulk regions, and EWCS in this phase is zero. Unlike mutual information, however, the transition of EWCS from positive values to zero is discontinuous.  

We follow the perturbation theory analysis outlined in the previous to calculate the EWCS for sub-region of very small size $\left(\ell \ll r_H\right)$. For simplicity, we consider a symmetric configuration where both $A$ and $B$ have the same width $\ell$. From the symmetry of figure \ref{fig_EWCS_setup}, one concludes that the minimal cross-section of the entanglement wedge is the bulk hypersurface $x\left(r\right)=\mathrm{constant}$. Its area is given by the integral
\begin{equation} \label{area_func_EWCS}
	\mathcal{A}_{\Sigma} = 2 L L_{\xi} \int_{r_{\ast}\left(h\right)}^{r_{\ast}\left(h+2\ell\right)} \frac{\alpha(r) H(r) K(r)^\frac{1}{3}}{r^2 \sqrt{f\left(r\right)}}\,dr ,
\end{equation}
which itself follows from equation \eqref{area_func} after we put $x(r)=\text{constant}$. Note that the limits of integration are the two turning-points of the two extremal area hypersurfaces which bound the entanglement wedge. We can also write the area integral as
\begin{align} \label{area_func_EWCS}
	\mathcal{A}_{\Sigma} = &2 L L_{\xi} \left[ \int_{\epsilon}^{r_{\ast}\left(h+2\ell\right)} \frac{\alpha(r) H(r) K(r)^\frac{1}{3}}{r^2 \sqrt{f\left(r\right)}}\,dr - \int_{\epsilon}^{r_{\ast}\left(h\right)} \frac{\alpha(r) H(r) K(r)^\frac{1}{3}}{r^2 \sqrt{f\left(r\right)}}\,dr \right],
\end{align}
which is easier to compute. The two integrals in the square brackets above differ only in their upper limits. Let us take one such integral and perform a series expansion in the small parameter $\frac{r_{\ast}}{r_H}$, we obtain
\begin{align*}
    \int_{\epsilon}^{r_{\ast}} \frac{\alpha(r) H(r) K(r)^\frac{1}{3}}{r^2 \sqrt{f\left(r\right)}}\,dr =\; & \frac{1}{\epsilon} - \frac{1}{r_{\ast}} \left( 1 + \frac{1}{2} \frac{r_{\ast}^2}{r_H^2} \left( q^2 - \beta^2 \left(1 + q^2\right)^3 \right) \right).
\end{align*}
Finally we use the perturbative formula for the turning-point from equation \eqref{turnpt_exp} to express the above as
\begin{align}
    \int_{\epsilon}^{r_{\ast}} \frac{\alpha(r) H(r) K(r)^\frac{1}{3}}{r^2 \sqrt{f\left(r\right)}}\,dr =\; & \frac{1}{\epsilon} - \frac{2 b_0}{\ell} \left(1 - \frac{1}{16} \frac{\ell^2}{r_H^2} \beta^2\left(1+q^2\right)^3 \left(\frac{3}{b_0^2} - \frac{b_1}{b_0^3} \right) \right),
\end{align}
where $b_0$ and $b_1$ are the numerical constants defined in equation \eqref{constant_defn}. Hence, the EWCS in the small strip-width approximation can be expressed as
\begin{equation}
    %
    \frac{4 G_{N}^{(5)} \times E_{W}\left(A, B \right)}{2 L L_{\xi}} = \frac{2 b_0}{\ell} \left( \left( \frac{1}{x} - \frac{1}{x-2} \right) - \frac{1}{8} \frac{\ell^2}{r_H^2} \beta^2\left(1+q^2\right)^3 \left(\frac{3}{b_0^2} - \frac{b_1}{b_0^3} \right) \right),
    %
\end{equation}
as before, we let $\frac{h}{\ell}=x$ denote the ratio of the separation between $A$ and $B$ two the width of each strip.
\begin{figure}[t]
    \begin{subfigure}{0.48\textwidth}
        \centering
        \includegraphics[width=\textwidth]{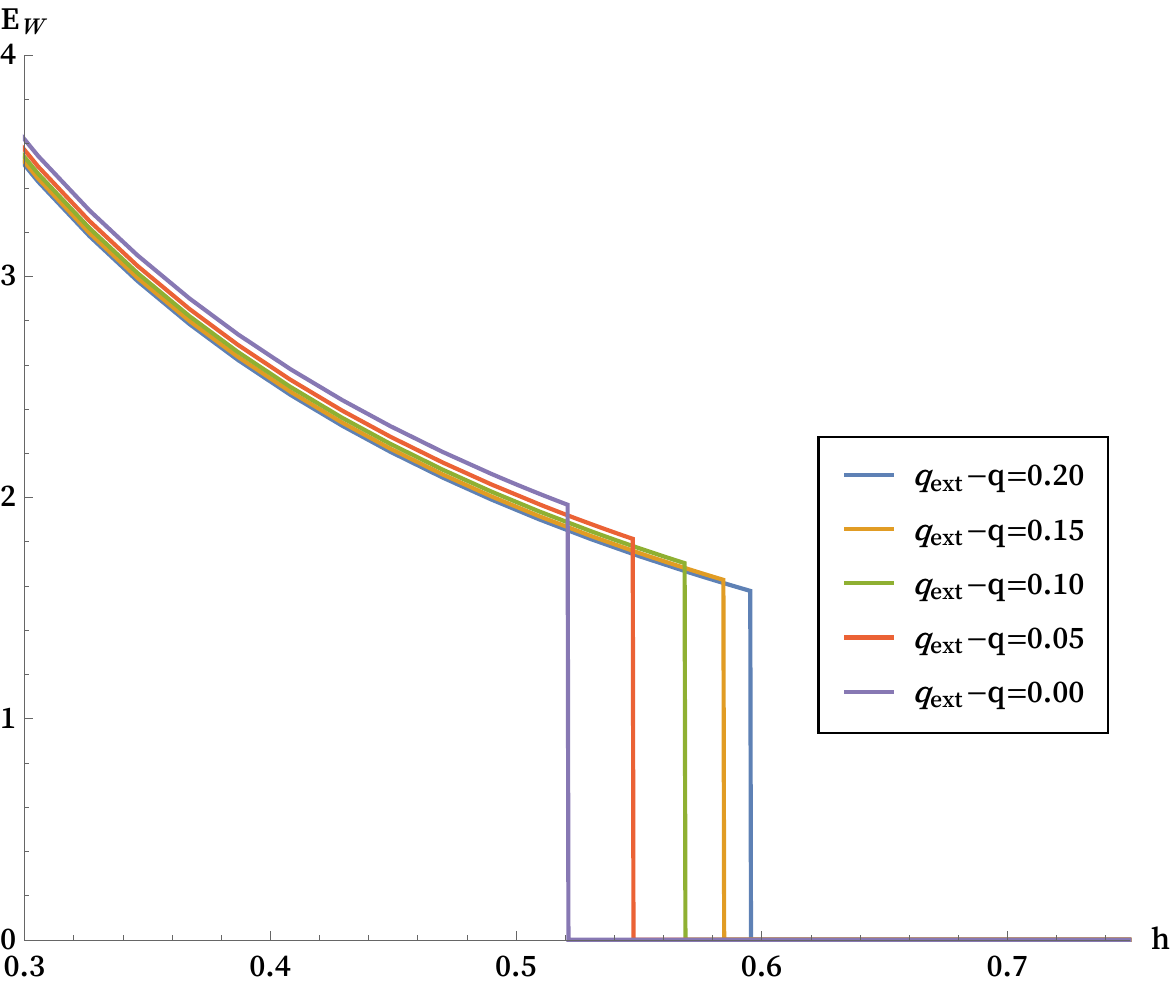}
        \caption{$\beta=1.0$}
    \end{subfigure}
    \hfill
    \begin{subfigure}{0.48\textwidth}
        \centering
        \includegraphics[width=\textwidth]{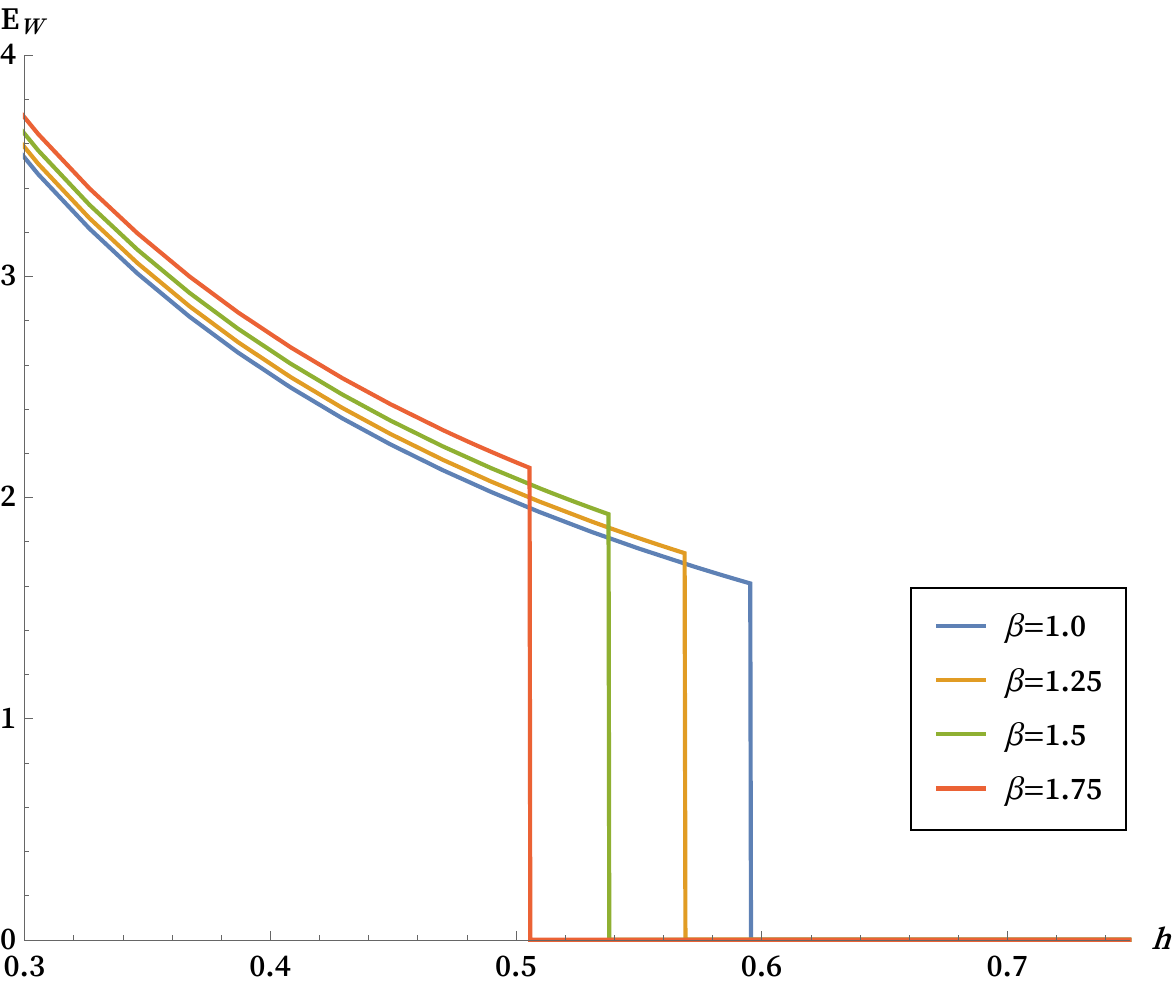}
        \caption{$q=0.30$}
    \end{subfigure}
    \caption{Dependence of entanglement wedge cross-section $E_W\left(A, B\right)$ on the separation $h$ between $A$ and $B$: (a) for different charges $q$, and (b) for different $\beta$ values. In both cases $r_H = 2\sqrt{2}$, and $L=L_{\xi}=\ell=1$. Plots are drawn in $4G_N^{(5)}=1$ unit.}
    \label{fig_EWCS}
\end{figure}
We study the dependence of the entanglement wedge cross-section on the two parameters $q$ and $\beta$ that characterize the Schro\"{o}dinger black hole. We have already figured out from our study of mutual information in the previous section that the `disentangling' phase transition occurs for smaller and smaller separation as both these parameters are increased. From figure \ref{fig_EWCS} we observe that when the charge $q$ is increased from below towards its extremal value $q_{\text{ext}}$, EWCS in the non-zero phase increases. Same conclusion holds for the other parameter $\beta$. Thus EWCS behaves in a manner qualitatively similar to holographic entanglement entropy, but different than mutual information with respect to change in electric charge $q$. An opposite conclusion is drawn as far as the dependence on $\beta$ is concerned; because HEE and mutual information both decrease with increasing $\beta$, while the EWCS increases. This is not at all surprising because the three quantities after all measure different kinds of correlation between $A$ and $B$, and it is expected that they will differ from one another here and there.

\section{Conclusion} \label{conclusions}

In this paper we used gauge/gravity duality to study certain quantum information theoretic measures in a $\left(4+1\right)$ dimensional charged Schr\"{o}dinger black hole geometry dual to a $\left(2+1\right)$ dimensional non-relativistic CFT state at finite temperature and with non-zero electric charge. The spacetime geometry is obtained from five-dimensional Reissner-Nordstr\"{o}m AdS black hole solution of type-IIB supergravity by a null Melvin twist. We studied holographic entanglement entropy, mutual information, and entanglement wedge cross-section for strip subsystems in this geometry. In particular, we discussed how these quantities depend on the two parameters which characterize the spacetime, namely the charge $q$, and the Melvinization parameter $\beta$. For holographic entanglement entropy we used both numerical methods and an approximate analytical computation for very small strip-width, while for the other two quantities we could not do a numerical analysis since the spacetime metric was too complicated to be controllable by the numerical methods we were following. However, given the very good matching of our numerical and perturbative calculations of entanglement entropy in the restricted parameter regime, we expect no qualitative difference for the other cases as well. Our work also serves as a check for the proposal of \cite{Kim:2012nb} to adopt the Ryu-Takayanagi formula for codimension two Schr\"{o}dinger holography, which we successfully apply in the course of our calculation.

It would be interesting to study directly the non-relativistic CFT, and check if the results commensurate with our holographic analysis. Some recent works have already studied entanglement of Schro\"{o}dinger field theory in the zero-temperature state \cite{Mintchev:2022xqh, Mintchev:2022yuo}. We focused on the strip geometry because of its simplicity, and it would be interesting to extend the analysis for compact entangling regions. Qualitatively, though, the entanglement measures should exhibit similar dependence on the characteristic length-scale associated with the subregion of choice. Another potential future direction is to study entanglement in a dynamical Schro\"{o}dinger black hole, generalizing the work of \cite{Roychowdhury:2016wca} for finite charge.

\bibliographystyle{JHEP}
\bibliography{references}

\end{document}